\documentclass[reprint,amsmath,amssymb,aps,
nofootinbib,floatfix]{revtex4-2}
\usepackage{graphicx}
\usepackage{dcolumn}
\usepackage{bm}
\usepackage{braket}
\usepackage{comment}
\usepackage[colorlinks=true, linkcolor=blue,
urlcolor=blue,citecolor=blue]{hyperref}
\usepackage{soul}

\newcommand{\di}{i} 

\begin{document}

\title{Quantum fluctuations induce collective multiphonons in finite Fermi liquids}

\author{Petar Marevi\'c}
\affiliation{Centre Borelli, ENS Paris-Saclay, 
Universit\'e Paris-Saclay, 91190 Gif-sur-Yvette, 
France}

\affiliation{Department of Physics, 
Faculty of Science, 
University of Zagreb, HR-10000 Zagreb, Croatia}
\email{pmarevic@phy.hr}

\author{David Regnier}
\affiliation{CEA, DAM, DIF, 91297 Arpajon, France}
\affiliation{Universit\'e Paris-Saclay, CEA, Laboratoire Mati\`ere en Conditions Extr\^emes, 91680 Bruy\`eres-le-Ch\^atel, France}

\author{Denis Lacroix } 
\affiliation{Universit\'e Paris-Saclay, CNRS/IN2P3, IJCLab, 91405 Orsay, France}

\begin{abstract}
We show that collective multiphonon states in atomic nuclei emerge at high excitation energies when 
quantum fluctuations in the collective
space are 
included beyond the independent-particle approximation. 
The quadrupole response of a nucleus is studied using an extension of the nuclear time-dependent density-functional theory 
that mixes several
many-body trajectories. While a single trajectory can account for the excitation of the first collective quantum, the second and the third quanta emerge due to the interference
between trajectories.
The collective spectrum, found as nearly harmonic, is in excellent agreement with the   
experimentally observed three quanta of the isoscalar giant quadrupole resonance in $^{40}$Ca. This study offers guidance 
for multiphonon searches in other self-bound systems and demonstrates the resistance
to internal excitation 
of finite Fermi liquids. 
\end{abstract}

\date{\today}

\maketitle

\section{Introduction}

The response to an
external perturbation 
is a powerful and versatile tool for studying properties of quantum many-body
systems \cite{pines2018}.
A recurrent feature of these
systems is the capacity of particles  
to self-organize into a collective motion,
leading to the occurrence of phonons in
quantum solids \cite{lipparini2008,leggett2006} and plasmons in electron gases \cite{dinh2018}. Such collectivity survives even in systems with 
a much smaller number of particles,
like in metallic clusters where the electronic
clouds exhibit collective motion with respect 
to the ions \cite{kresin1992}, or in self-bound
finite systems like atomic nuclei where
the so-called giant resonances (GRs)
have been studied for decades
\cite{rowe2010,harakeh2006,bortignon2019}.
A convenient framework for understanding
the dynamics of quantum liquids
is the time-dependent density-functional
theory (TDDFT) \cite{lacroix2004,simenel2010,simenel2012,bulgac2013,casida12,marques2012,schunck2019}.
In the linear-response regime of TDDFT \cite{ring2004,marques2012},
the GRs represent the first harmonic quantum
of quasibosonic excitations.
Based on this picture,
it is expected that higher oscillatory
quanta, the so-called multiphonon
states,
should exist as well \cite{chomaz1995,aumann1998}.
After several decades of
research, the existence of two-phonon
GRs in several nuclei
has been
established using multiple
experimental techniques,
including the
heavy-ion inelastic scattering at intermediate
or relativistic energies
and the
pion double charge exchange 
reactions
\cite{mordechai1988a,mordechai1988b,schmidt1993,
ritman1993,
mordechai1996,boretzky1996,boretzky2003,scarpaci1993,
scarpaci1997}.
An exciting case is
the $^{40}$Ca nucleus where, in addition
to a two-phonon state \cite{scarpaci1993,scarpaci1997},
an experimental signature of
a three-phonon state was
reported as well
\cite{fallot2006}. Such multiphonon states built from the collective motion 
of valence electrons have also been probed in metallic clusters 
\cite{brechignac1990,brechignac1992,gambacurta2009}.
Despite these efforts, the 
experimental signatures of multiphonon
states in finite quantum systems are scarce, and many questions remain open regarding their stability against internal disorder, particle emissions,
or even how they emerge from a microscopic picture.

A difficulty for microscopic modeling of
multiphonon excitations
is the fact that the TDDFT motion reduces to 
a quasiclassical evolution of
the one-body degrees of freedom \cite{blaizot1986,chomaz1995}, leading
to a large underestimation of quantum
fluctuations in the collective space.
Even though the oscillations of TDDFT densities
may contain some signatures of a second phonon
\cite{ring1996},
the small amplitude limit of TDDFT 
gives only an access to the first quantum of 
collective state.
Higher excitation quanta can then be generated
by an {\it ad hoc} introduction of phonon degrees
of freedom through,
for example, the boson mapping
method \cite{andres2001,lanza2006}
or the quasiparticle-phonon model
\cite{soloviev1992,bertulani1999}.
Alternatively, a model can be explicitly
requantized through, for example, a
semiclassical quantization of 
periodic orbitals \cite{negele1982,chomaz1995}
or the fully quantum configuration mixing
method \cite{hill1953,griffin1957,goeke1982,reinhard1987}. Today,
the latter approaches typically employ bases of time-independent many-body states~\cite{verriere2020} chosen on the adiabatic energy landscape, although a few attempts leveraging time-dependent bases have been recently made within schematic models \cite{regnier2019,hasegawa2020}. As demonstrated
by TDDFT, collective excitations 
can hardly be reconciled with the adiabatic assumption unless an enormous number of states is used.    

In this work, the method
discussed in Refs.~\cite{reinhard1983,regnier2019}
is extended and
applied to realistic
modeling of nuclear dynamics.
The TDDFT evolution is requantized through the mixing of several
TDDFT trajectories.
By applying the model to the
case of GRs in the $^{40}$Ca
nucleus,
it is demonstrated that the 
collective multiphonon excitations emerge naturally in the collective response and
their origin is attributed to the influence of quantum fluctuations in the collective space.

\section{Theoretical framework}

The basic ingredient
of our method is the nuclear TDDFT,
which is considered the most advanced microscopic approach to
small and large amplitude
nuclear motion beyond the adiabatic limit \cite{simenel2012,bulgac2019}. While a single TDDFT 
trajectory is rather predictive for the motion of one-body observables, it largely underestimates quantum 
fluctuations in the collective
space that are beyond the independent-particle approximation. Significant efforts have been made to 
include fluctuations via phase-space methods \cite{lacroix2014,tanimura2017} using independent TDDFT trajectories. 
Even though such approaches
account for quantum
fluctuations, a
genuine quantization in 
the collective space 
is required to
describe the quantum
interference between 
trajectories.

\begin{figure}
    \centering
    \includegraphics[width=0.45\textwidth]{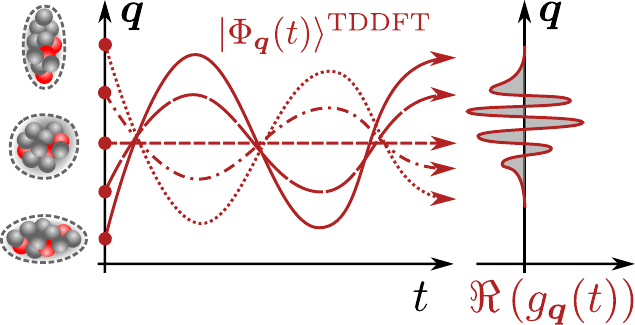}
    \caption{Illustration of the MC-TDDFT approach to the quantum many-body dynamics. The nuclear
    wave function is a weighted, time-dependent superposition of non-orthogonal Slater determinants $\ket{\Phi_{\bm{q}}(t)}$, each following its own TDDFT trajectory. The
    complex collective wave function $g_{\bm{q}}(t)$ accounts for quantum interference between trajectories.
    }
    \label{fig:collective_dyn}
\end{figure}

Based on the
well-known static
generator coordinate method (GCM) \cite{hill1953,griffin1957},
a natural way to re-quantize the collective nuclear motion is to write the many-body wave function
as
\begin{equation}
\ket{\Psi(t)} = \sum_{\bm{q}} f_{\bm{q}}(t)
\ket{\Phi_{\bm{q}}(t)},
\label{eq:MCTDHF_State}
\end{equation}
where $\{ \ket{\Phi_{\bm{q}}(t)} \}$
represents a set of time-dependent Slater determinants labeled by a
vector of collective
coordinates $\bm{q}$.
The many-body state of~\eqref{eq:MCTDHF_State} is 
formally equivalent
to the one used in quantum chemistry within the
so-called
multiconfigurational time-dependent
Hartree-Fock (MC-TDHF) framework \cite{meyer1990,beck2000,meyer2009}.
In MC-TDHF, the trajectories
are orthogonal to each other
and their evolution follows
a coupled set of TDHF equations.
However, two distinctions in the
nuclear case need to be made:
(i) the starting point of our calculation
is the nuclear TDDFT based on phenomenological
nuclear energy density functionals (NEDFs) \cite{schunck2019,ring2004},
(ii) the set of 
generating states 
$\{ \ket{\Phi_{\bm{q}}(t)} \}$ is
typically non-orthogonal, rendering
the practical implementation
significantly
more complex. For these reasons,
the nuclear model presented here
will be referred
to as the multiconfigurational
TDDFT (MC-TDDFT).

To overcome the difficulties listed
above that are particular
to the nuclear case, and
following the work of Refs.~\cite{reinhard1983,regnier2019}, we assume here that each state in Eq.~\eqref{eq:MCTDHF_State}
follows its own TDDFT trajectory,
that is
\begin{equation}
\di \hbar \dot{\rho}_{\bm{q}}(t) = \Big[h[\rho_{\bm{q}}(t)],\rho_{\bm{q}}(t)\Big],
\end{equation}
where $\rho_{\bm{q}}(t)$ is the
one-body density matrix corresponding
to $\ket{\Phi_{\bm{q}}(t)}$
and $h[\rho_{\bm{q}}(t)]$ is the single-particle
Hamiltonian deduced from an effective interaction \cite{schunck2019,ring2004}.
The mixing coefficients 
$f_{\bm{q}}(t)$, 
that account for quantum interference
between trajectories, are determined through
the Dirac-Frenkel variational
principle which
requires that the action
\begin{equation}
S = \int_{t_0}^{t_1}
\,dt \braket{\Psi(t)|
\hat{H} - \di \hbar \partial_t
| \Psi(t)}
\end{equation}
is stationary with respect
to variation of the state \eqref{eq:MCTDHF_State}.
By treating the mixing coefficients as
variational
parameters we obtain 
\begin{equation}
\di \hbar \dot{g} = 
\Big[
\mathcal{N}^{-1/2}(\mathcal{H} - \mathcal{D})
\mathcal{N}^{-1/2}
+ \di \hbar \dot{\mathcal{N}}^{1/2}
\mathcal{N}^{-1/2} \Big] g.
\label{eq:g_Evolution}
\end{equation}
Here, $\mathcal{N}_{\bm{q} \bm{q'}}(t) = 
\braket{\Phi_{\bm{q}}(t)|\Phi_{\bm{q'}}(t)}$
corresponds to the overlap kernel,
while
$\mathcal{H}_{\bm{q} \bm{q'}}(t) =
\braket{\Phi_{\bm{q}}(t)|\hat{H}|\Phi_{\bm{q'}}(t)}$
is the Hamiltonian kernel whose
density-dependent part
is calculated with the
average density prescription
\cite{schunck2019}.
Finally, the derivative kernel reads
$\mathcal{D}_{\bm{q} \bm{q'}}(t) =
\braket{\Phi_{\bm{q}}(t)|\di \hbar \partial_t|\Phi_{\bm{q'}}(t)}$.
The normalized
collective wave function,
$g_{\bm{q}}(t) = 
\sum_{\bm{q'}} \mathcal{N}^{1/2}_{\bm{q} \bm{q'}}(t) f_{\bm{q'}}(t)$,
are determined by the numerical
solution of Eq.~\eqref{eq:g_Evolution}
and can be used to calculate
the expectation value
of any observable.

The present framework is 
similar in spirit
to the time-dependent generator
coordinate method
(TDGCM)
\cite{verriere2020} whose
adiabatic realization is particularly popular in fission studies
\cite{schunck2022}. However,
a major difference is
that MC-TDDFT
automatically
incorporates the
nonadiabatic effects by
considering TDDFT trajectories.
Consequently, we expect 
a drastic reduction in the number of
generating states as compared 
with the adiabatic TDGCM, 
even if relatively 
high internal excitation energies
are considered. 
Figure \ref{fig:collective_dyn}
illustrates the essential concepts
of the MC-TDDFT approach
to the quantum
many-body dynamics.

\section{Results}

TDDFT calculations are performed using
a newly developed code
based on the finite element
method \cite{zienkiewicz2013,mfem}.
The code was 
benchmarked against the
Sky3D code
\cite{schuetrumpf2018}; 
a detailed performance analysis  will
be reported elsewhere.
Dynamics of the $^{40}$Ca nucleus is simulated in
a three-dimensional box of
length $L = 24$ fm,
with a regular mesh of $14$ cells
in each
spatial direction and a
finite element basis of
third-order polynomials.
We use the
SLy4d NEDF \cite{kim1997} which
is particularly well suited
for nuclear dynamics studies.
Starting from the NEDF ground state 
$\ket{\Phi_1(0)}$, different configurations can be 
obtained by applying 
the isoscalar quadrupole boost operator
$\exp{(\di \lambda \hat{Q}_{20})}$,
where $\hat{Q}_{20}$
is the quadrupole operator
and $\lambda$ is the boost amplitude
\cite{simenel2012,scamps2013}.
For this study, we choose a minimal
mixed state 
of the form
\eqref{eq:MCTDHF_State}
with $\bm{q} = 1, 2, 3$,
where $\ket{\Phi_2(0)}$ and $\ket{\Phi_3(0)}$
are quadrupole-boosted states with the excitation energies
of about $0.25$
and $0.50$ MeV, respectively. These
configurations are chosen so that the basis
of generating states
is not overly linearly dependent.
To study the isoscalar 
giant quadrupole resonance (GQR), the entire
mixture is perturbed with
a common boost of $\lambda = 5.7 \times 10^{-3}$
fm$^{-2}$ and
left
to evolve in time. The
time step of $\Delta t = 5 \times
10^{-25}$~s was verified
to provide well-converged
results. The mixing coefficients
were set to $f_1(0) = 1$ and $f_2(0) = f_3(0) = 0$
so that the initial state corresponds to a single Slater determinant 
while the interference
with other trajectories
kicks in during time evolution. 

\begin{figure}
\includegraphics[width=0.49\textwidth]{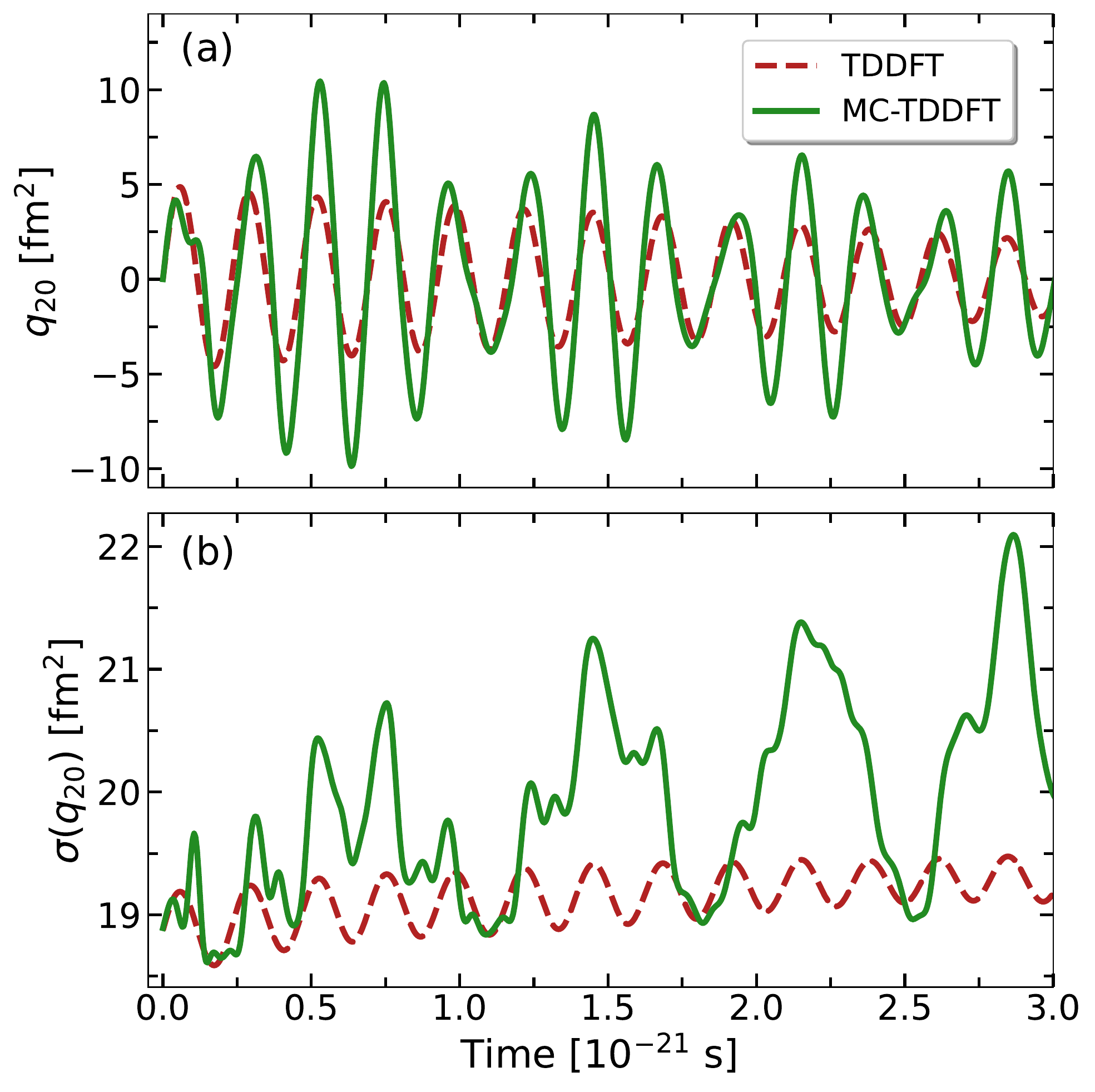}
\caption{(a) The response of the $^{40}$Ca nucleus to an isoscalar
quadrupole perturbation,
obtained for a single TDDFT 
trajectory (red dashed line) and for the MC-TDDFT state (green solid line).
(b) The corresponding fluctuations,
$\sigma(q_{20}) =
\sqrt{
\braket{\hat Q^2_{20}}
- \braket{\hat Q_{20}}^2}$.   
The inclusion of quantum fluctuations beyond
the independent-particle approximation gives rise to the onset
of multiple frequencies.
}
\label{fig:q20}
\end{figure}

Figure \ref{fig:q20}(a) shows the response
of the $^{40}$Ca nucleus 
to the isoscalar
quadrupole boost, either assuming no interference (a single TDDFT trajectory)
or when the interference
is taken into account (the MC-TDDFT case).  
Given that the oscillation period
of GRs is typically of the order of 
$10^{-22}$~s \cite{chomaz1995},
the nuclear dynamics is simulated up to $t = 3 \times 10^{-21}$~s.
The TDDFT curve exhibits almost harmonic oscillations with a single frequency, consistently
with what is usually observed within TDDFT
when
the Landau damping effect
is absent \cite{pines2018,harakeh2006,lacroix2004}. Note that the other two TDDFT trajectories (not shown)  
yield similar oscillations
with slightly larger amplitudes.

On the other hand, the $q_{20}(t)$ evolution of the MC-TDDFT state
is markedly more complex,
with the emergence of both
lower and higher new frequencies.
Such beating of collective modes is seen
even more clearly
in the corresponding fluctuations,
shown in Fig.~\ref{fig:q20}(b).
While fluctuations in 
the TDDFT case have rather
modest amplitudes and
oscillate with the same frequency as $q_{20}(t)$,
the MC-TDDFT fluctuations exhibit significantly
larger
amplitudes and include the beating of
various frequencies. 

\begin{figure*}
\includegraphics[width=0.98\textwidth]{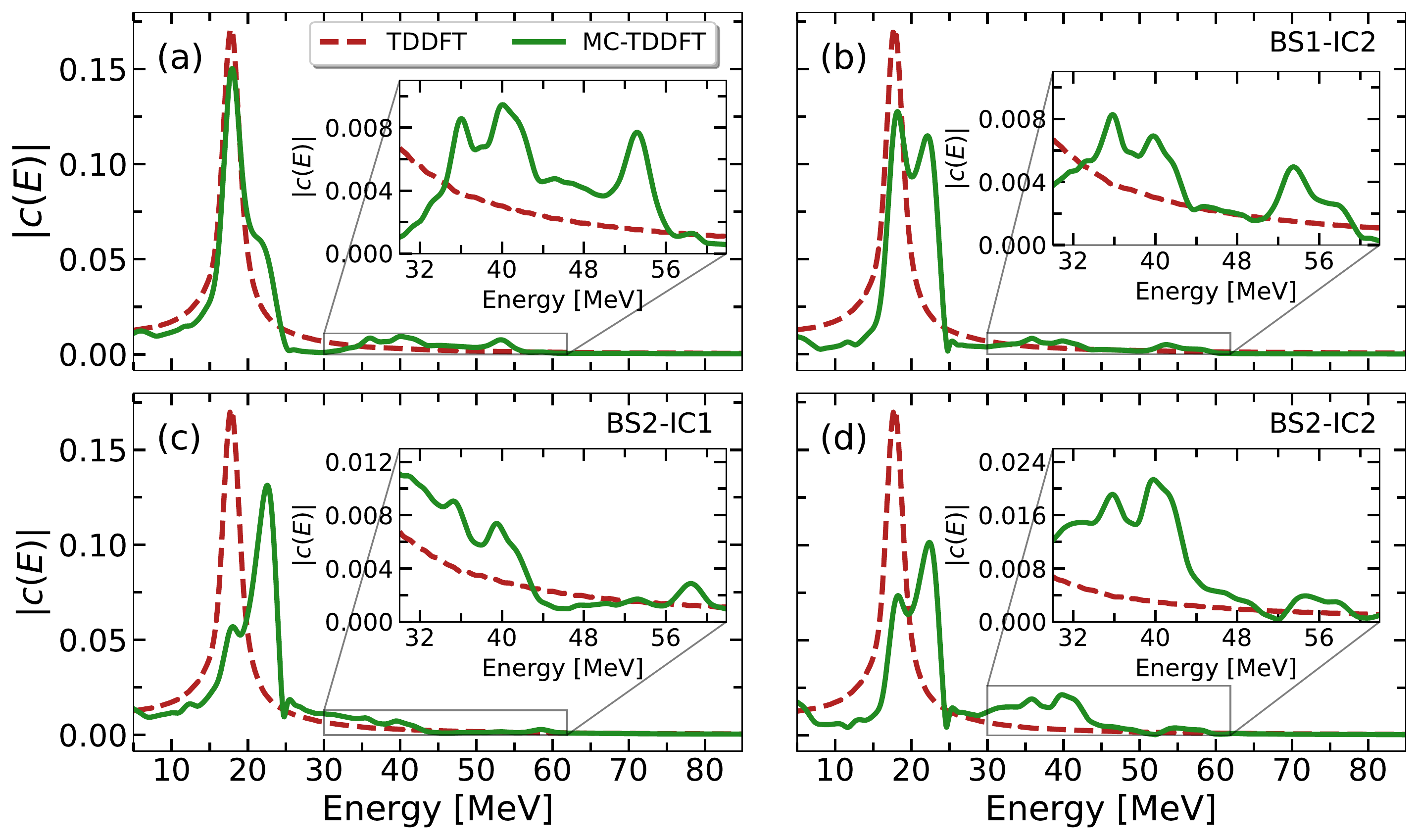}
\caption{(a): The excitation spectrum
of the isoscalar giant quadrupole resonance
in $^{40}$Ca. The TDDFT approach (red dashed line) yields
one frequency, in agreement
with the classical oscillator picture.
Note that the width of the peak is partially
induced by the value of $\Gamma_0$ in Eq.~\eqref{eq:ce}.
The
MC-TDDFT spectrum (red solid line) is
fragmented
around the main GR peak, and new high-energy components appear
at twice and three times the energy of
the main peak (see inset).
Panels
(b), (c), and (d) show
the same excitation
spectrum for different combinations
of basis states
and initial conditions defined in the
text:
BS1-IC2, BS2-IC1, and
BS2-IC2, respectively.
The appearance of multiphonon structures
and their energies are robust with respect to the change of basis
states and initial
conditions.
}
\label{fig:Strength}
\end{figure*}

A quantitative analysis of the quadrupole response can be made using the Fourier transform method.
The excitation spectrum of GRs
is calculated as
\begin{eqnarray}
\left| c(E) \right| &=& \left| 
\int_0^\infty \,dt 
q_{20}(t) f(\Gamma_0, t)
\exp\left(\di \frac{E}{\hbar} t \right)
 \right|,
\label{eq:ce}
\end{eqnarray}
where the damping function
$f(\Gamma_0, t)$ is a conventional
way to account
for the fact that time evolution
can be simulated only up 
to finite times. In practice,
we used a Gaussian damping which
ensures a good resolution
of individual peaks. The
form of the Gaussian is such
that it is equivalent to the
standard exponential decay
with the
same $\Gamma_0 = 1.5$ MeV parameter.

In Figure \ref{fig:Strength}(a),
we show the normalized $|c(E)|$ 
obtained from the TDDFT and
the MC-TDDFT quadrupole responses.
The TDDFT curve exhibits
one peak centered
at about $E_{\rm{TDDFT}} = 17.8$ MeV,
in agreement with the classical
oscillator picture. 
Of course, the peak energy
may slightly differ for other
Skyrme parametrizations.
Note that the other two
TDDFT trajectories used in MC-TDDFT 
yield essentially indistinguishable curves,
as expected in the
small oscillations limit.
As could be anticipated from Fig.~\ref{fig:q20}(a), the
MC-TDDFT once again 
yields a markedly richer structure.
In particular, the main GR peak is
slightly shifted and it gets fragmented with 
another peak at 
about $4$ MeV
higher energy. 
Experimentally,
the existence
of the isoscalar GQR
in $^{40}$Ca in the 
$E_{\rm{GR}} \approx 18$ MeV region
has long been established
\cite{torizuka1975,marty1975,
youngblood1976,youngblood1977,
moalem1977,yamagata1978,
arvieux1979,lui1981,
youngblood2001}.
Moreover, a fragmentation of the
main peak with a secondary
contribution at about $3.5$ 
MeV lower energy
was also observed
\cite{moalem1977,scarpaci1993}.
Theoretical models based on the
random-phase approximation (RPA)
are able to predict a fragmentation 
similar to the experimental one only 
when extended with
complex
damping mechanisms stemming from
the 
nucleon-nucleon collisions and 
coupling to low-lying states \cite{lacroix2004}.
Even though the present calculation
does not capture fine details
of this fragmentation,
it is interesting to note
that the magnitude of splitting
is roughly reproduced based on
an entirely different framework.
Note that the present calculations
consider very few configurations
compared with the extended RPA
framework. It would therefore be interesting
to explore how this fragmentation
evolves as more collective degrees
of freedom are taken into account.

Most interestingly, the new feature in Fig.~\ref{fig:Strength}(a)
are the structures in the 
$30$ MeV $\leq E \leq$ $60$ MeV
region:
two peaks at roughly
twice and three times
the energy of the
first fragmented peak. 
We verified the robustness
of these structures
with respect to the change of initial conditions and the choice of TDDFT
trajectories.
In particular, in addition
to the set
of basis states
(referred to as BS1) and the 
set of initial
conditions 
(referred to
as IC1) discussed above,
we considered two more
sets. For the second
set of basis states (BS2),
we start
by quadrupole-boosting the NEDF
ground state
$\ket{\Phi_1(0)}$
by about $0.5$ MeV. The
three states of the BS2 set
then correspond to
(i) the boosted state at $t = 0$, (ii) the boosted
state at time $\tau$ when
maximal deformation 
$q_{20}(\tau)$
is reached for the first time, (iii) and the boosted state at time 
$\tau/2$.
Furthermore, the second
set of initial conditions
(IC2) is determined
by diagonalizing the collective
Hamiltonian at $t=0$,
rendering the initial MC-TDDFT state equal
to the actual ground state.
The excitation spectrum for 
BS1-IC2, BS2-IC1, BS2-IC2
combinations is shown in
panels \ref{fig:Strength}(b)-\ref{fig:Strength}(d),
respectively.
While relative height of peaks
depends on
the way the initial
state is prepared and the
choice of basis states,
the appearance of peaks
and their energies
are very robust. Additionally,
we verified that
the structures persist when another 
Skyrme effective interaction is used.

\begin{figure}
\includegraphics[width=0.49\textwidth]{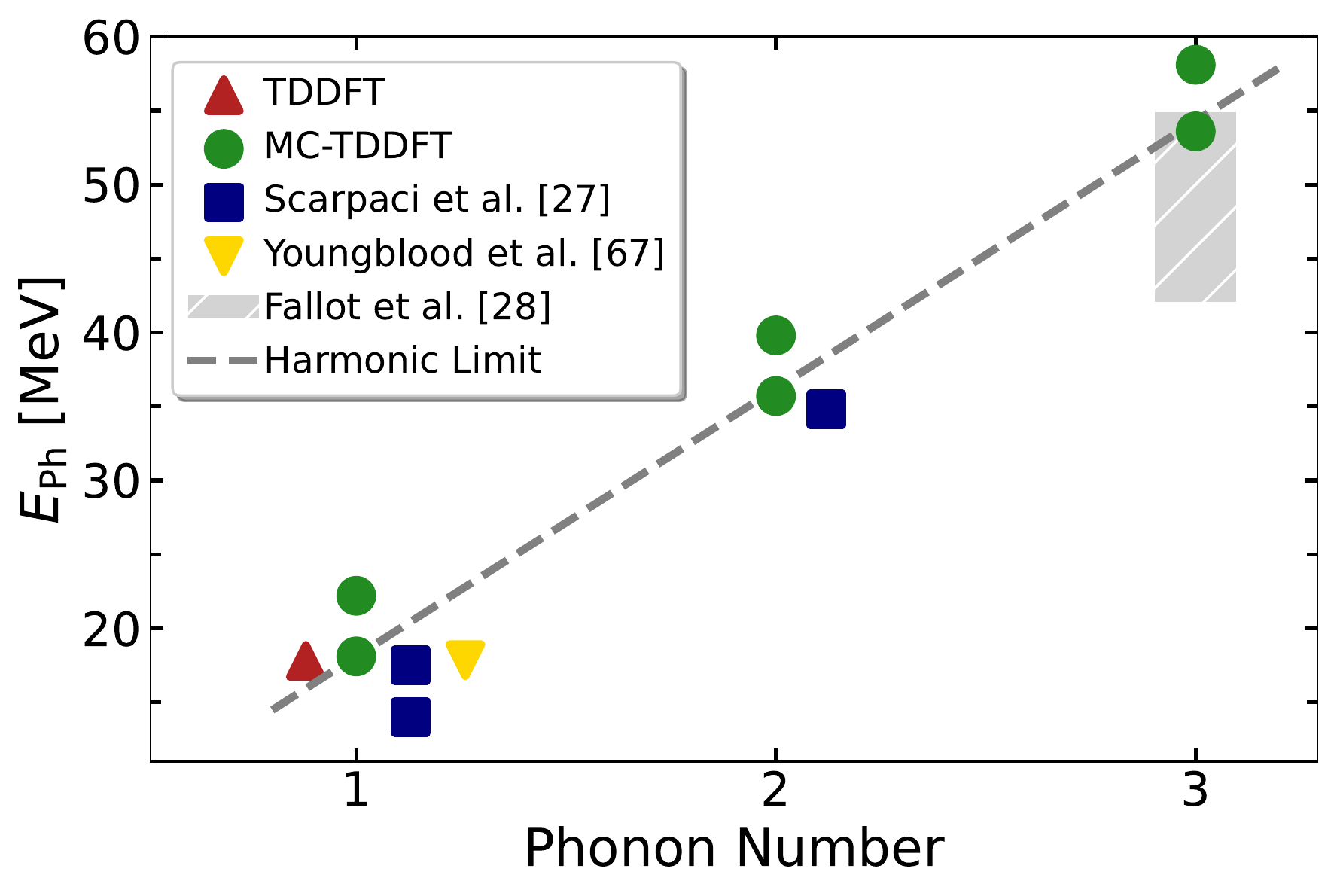}
\caption{The mean excitation energies
of multiphonon states obtained
in MC-TDDFT.
The first and the second phonon
are in excellent agreement
with experimental data 
\cite{youngblood2001,scarpaci1997}.
We predict the third photon
within the energy range
observed in Ref.~\cite{fallot2006}
(the shaded gray area).
The excitation pattern
of the 
main peak
closely follows that
of the
harmonic oscillator
(the dashed
gray line).
Note that certain symbols are shifted horizontally
to avoid overlaps.}
\label{fig:Energies}
\end{figure}

The fact that these high-energy structures
exhibit the same splitting as the
original GR represents a strong indication
that they correspond to the
second and the third quantum
of the same
excitation. 
Without making any {\it a priori} assumptions
on their existence,
MC-TDDFT predicts
the occurrence of
multiphonon states
on top of the main
GQR.
Figure \ref{fig:Energies} summarizes the calculated energies
of multiphonon states,
alongside the representative
experimental data. 
The $1\sigma$ uncertainties were
estimated from the four different
combinations of basis states
and initial conditions
defined above.
This estimate
does not include the uncertainty
due to the nuclear interaction
and other possible sources.
The main one-phonon
peak is found
at $E_{\rm{ph}} =
(18.1 \pm 0.2)$ MeV,
in excellent
agreement with the bulk
of experimental
data.
In Refs.~\cite{scarpaci1993,scarpaci1997},
the
inelastic scattering of $^{40}$Ca on $^{40}$Ca 
at $50$ MeV/nucleon was measured in coincidence with protons
and a two-phonon state at
$(34.8 \pm 0.5)$ MeV was established.
This is very close to our value of 
$E_{2\rm{ph}} = (35.7 \pm 0.4)$ MeV.
Finally,
later inelastic-scattering experiments
have found a three-phonon structure
built on the isoscalar GQR
in the energy range of $42-55$ MeV
\cite{fallot2006}
but were not able to precisely
determine its energy.
Another experiment hinted at
a possibility of a three-phonon giant dipole resonance
in $^{136}$Xe \cite{schmidt1993}, but
the corresponding cross section was
low and the authors were not
able to exclude instrumental effects.
To the best of our knowledge,
these are the only
experimental observations
of the kind in atomic nuclei.
The MC-TDDFT model confirms
the existence of a three-phonon isoscalar GQR state
in $^{40}$Ca with an energy 
$E_{3\rm{ph}} = (53.6 \pm 0.7)$
MeV, within
the relevant experimental range. Besides this confirmation, our calculations also demonstrate
the survival of such excitation for a sufficiently long time despite the high internal excitation of the system.  

We note that present
calculations
do not predict
any additional structures
up to $100$ MeV,
including the region
of a hypothetical fourth phonon at
$E \approx 72$ MeV.
The obtained results support the 
quasibosonic
picture of multiphonon
states as essentially
harmonic excitations \cite{chomaz1995}.
The presence of anharmonicities is 
minor (see the very slight
departure from the dashed line in 
Fig.~\ref{fig:Energies};
based on the reported excitation
energies, we estimate it
to be of the order of $2\%$.
This result is nontrivial because
(i) no harmonic approximation was 
explicitly made
in our calculations 
and (ii) it was argued
that some mesoscopic systems, such as 
metallic clusters, can exhibit extremely
strong anharmonicities \cite{catara1993}.

\section{Conclusion}

We showed that multiphonon
collective states emerge naturally
within the new multiconfigurational
TDDFT
model that includes quantum fluctuations
in the collective space
beyond the independent-particle approximation. 
The experimentally observed multiphonon states
in $^{40}$Ca are obtained at nearly-harmonic excitation
energies, in excellent agreement with experiments
and without an \textit{ad hoc}
introduction of phonon degrees of freedom.
Beyond its relevance for
collective excitations,
the MC-TDDFT model and its extensions
can be used to study
a variety of nuclear phenomena. For example,
the inclusion of pairing correlations
would
be crucial for TDDFT studies of fission
where the lack of quantum
fluctuations prevents
the calculation of
quantities such as
mass yields or angular
momenta for the entire
range of observed fragmentations \cite{schunck2022}.
Finally, while we focused on atomic
nuclei, this study may offer
guidance for multiphonon
searches in other self-bound systems,
such as quantum droplets that were
recently formed in laboratory
\cite{baillie2017,bottcher2020,chomaz2023}.

\bibliography{bibliography}
\end{document}